# Atomic-scale order enables high thermal boundary conductance at $\beta$-Ga$_2$O$_3$/4H-SiC interfaces

Hongao Yang[1,6], Yongtao Yang[1,2,6], Yuanbin Liu[1,3,6], Tao Ding[1], Yang Shen[1], Jiawei Huang[4], Weigang Ma[1], Linfeng Fei[4], Zhenping Wu[2,*], Gábor Csányi[3,5,*] & Bingyang Cao[1,*]

[1]*Key Laboratory of Thermal Science and Power Engineering of Ministry of Education, Department of Engineering Mechanics, Tsinghua University, Beijing 100084, China*
[2]*State Key Laboratory of Information Photonic and Optical Communications & School of Physical Science and Technology, Beijing University of Posts and Telecommunications, Beijing 100876, China*
[3]*Engineering Laboratory, University of Cambridge, Cambridge CB2 1PZ, United Kingdom*
[4]*School of Physics and Materials Science, Jiangxi Provincial Key Laboratory of Photodetectors, Nanchang University, Nanchang, Jiangxi 330031, China*
[5]*Max Planck Institute for Polymer Research, Ackermannweg 10, 55128 Mainz, Germany*
[6]*These authors contributed equally: Hongao Yang, Yongtao Yang, Yuanbin Liu*

*E-mail: zhenpingwu@bupt.edu.cn; gc121@cam.ac.uk; caoby@tsinghua.edu.cn

# Abstract

Thermal boundary conductance (TBC) at dissimilar interfaces imposes a fundamental limit on electronic device performance, yet predicting and understanding heat transport across realistic, disordered boundaries remains elusive. Here, we develop a computational framework that combines machine-learned interatomic potentials with lattice dynamics to address the long-standing problem of how interfacial structure, from disordered to atomically sharp, affects thermal transport in the technologically important $\beta$-Ga$_2$O$_3$/4H-SiC heterostructure. By explicitly accounting for phonon wave–particle duality, we show that interfacial disorder introduces additional interfacial phonon modes that facilitate vibrational impedance matching between the two highly dissimilar crystals, yet it simultaneously disrupts interfacial phonon coherence and limits the potential heat-transport benefit. Our atomistic simulations further indicate that restoring atomic-scale order preserves coherence and yields markedly higher conductance. These insights motivate the controlled epitaxial growth of $\beta$-Ga$_2$O$_3$/4H-SiC heterostructures with systematically tuned interfacial order. Experimental measurements validate our predictions, achieving a record-high TBC of 231 MW m$^{-2}$ K$^{-1}$ at atomically sharp junctions. Beyond the immediate implications for $\beta$-Ga$_2$O$_3$-based power electronics, our results establish the preservation of interfacial phonon coherence as an effective strategy for mitigating thermal bottlenecks in mismatched systems.

# Introduction

As electronic devices are pushing toward higher power densities and aggressive downscaling[1,2], managing extreme heat flux (up to 3,000 W cm$^{-2}$)[2] has become a grand challenge for performance and reliability[3]. Ultrahigh-thermal-conductivity substrates, such as diamond (~2,000 W m$^{-1}$ K$^{-1}$)[4], cubic boron arsenide (~1,300 W m$^{-1}$ K$^{-1}$)[5–8], and $\theta$-phase tantalum nitride (~1,100 W m$^{-1}$ K$^{-1}$)[9,10], have been proposed to offer exceptional heat spreading. As bulk resistance diminishes, the thermal bottleneck shifts to buried interfaces, where low thermal boundary conductance (TBC) can nullify the benefit of even the most conductive substrate[11,12], leading to interfacial heat transport as the critical factor for effective thermal management[12–14].

Interfacial engineering offers a direct route to enhance thermal conductance by tuning interfacial bonding and coordination to promote phonon transmission[11,15,16]. Disordered interfaces are particularly compelling in this regard, not only because they are pervasive in practical devices as a result of processing steps such as oxidation, ion-beam treatments and surface passivation[17,18], but also because they may, counterintuitively, be leveraged to enhance interfacial heat transport. Recent studies suggest that such ultrathin disordered interphases can act as phonon bridges between vibrationally dissimilar crystals, strengthening interfacial coupling and opening additional energy-transfer pathways[19–21]. Yet this picture appears inconsistent with several experimental observations reporting suppressed interfacial heat transport in the presence of disorder[17,22,23]. Identifying how disorder affects interfacial heat transport, and which mechanisms dominate in realistic interphases, remains an underexplored but practically important problem for interface-enabled thermal management.

The accurate modeling and understanding of interfacial heat transport in realistic devices is, however, a formidable challenge as conventional theoretical approaches struggle to capture the phonon-scattering pathways reshaped by interfacial disorder and consequently tend to overestimate TBC[13]. Molecular dynamics (MD) simulations are limited by classical approximations that neglect both quantum coherence and Bose–Einstein statistics[24] as well as by force-field inaccuracies[25]. By contrast, analytical approaches such as the acoustic mismatch model (AMM) and diffuse mismatch model (DMM)[26] completely omits explicit interfacial structure and bonding. These limitations can manifest as substantial overestimates of thermal conductance. For instance, a MD study of Si/diamond interfaces with an amorphous interlayer predicted ~381 MW m$^{-2}$ K$^{-1}$ despite experimental values near ~64 MW m$^{-2}$ K$^{-1}$ [27], while DMM calculations for $β$-Ga$_2$O$_3$/diamond yielded ~312 MW m$^{-2}$ K$^{-1}$, again far above the experimental value of ~17 MW m$^{-2}$ K$^{-1}$ [28]. Such discrepancies between computations and experiments underscore the need for more rigorous approaches that explicitly incorporate realistic interfacial structure and an appropriate quantum description of heat transport across interfaces. Bridging this gap is not merely a matter of numerical accuracy; it is a prerequisite for moving from post hoc interpretation to predictive control of interfacial thermal transport.

Here, we show how interfacial disorder governs interfacial heat transport by integrating advanced machine-learned interatomic potentials (MLIPs) with a wave–particle hybrid model (WPHM) grounded in lattice dynamics (LD) theory[29], and how this framework can in turn guide the design and synthesis of high-TBC interfaces. The $β$-Ga$_2$O$_3$/4H-SiC interface is studied as a prototypical system. As an ultrawide-bandgap semiconductor, $β$-Ga$_2$O$_3$ is highly promising for next-generation high-power electronics but severely limited by self-heating due to its low intrinsic thermal conductivity (11–27 W m$^{-1}$ K$^{-1}$)[30]. In contrast, 4H-SiC serves as a high-thermal-conductivity substrate that provides an ideal heat-spreading pathway. This sharp thermal contrast renders interfacial heat flow the decisive factor governing device performance, making the mitigation of self-heating especially critical to realizing the technological potential of $β$-Ga$_2$O$_3$. Experimental reports have demonstrated that interfacial disorder exerts a significant influence on the interfacial heat transport of $β$-Ga$_2$O$_3$/4H-SiC interfaces, with measured values spanning 23–150 MW m$^{-2}$ K$^{-1}$ [31–35]. Yet, the microscopic origins of this behavior and the limit of interfacial heat transport in such systems remain elusive. Leveraging our MLIP–WPHM framework, we bridge this knowledge gap by directly linking atomic-scale disorder to phonon transmission and interfacial heat transport, and by identifying design principles that enable high-conductance interfaces. Guided by these insights, we experimentally synthesized a series of $β$-Ga$_2$O$_3$/4H-SiC interfaces with varying degrees of disorder through heteroepitaxial growth. The experimental measurements align with theoretical predictions, achieving a record-high TBC of 231 MW m$^{-2}$ K$^{-1}$ in the disorder-free limit. Our findings not only advance $β$-Ga$_2$O$_3$ device applications but also give insight into interfacial phonon transport and showcase machine learning potentials as powerful tools for helping to engineer real-world complex heterostructured materials.

# Results

**Phonon scattering at *β*-Ga$_2$O$_3$/4H-SiC interfaces**

To model the phonon transport at *β*-Ga$_2$O$_3$/4H-SiC interfaces with *ab initio* accuracy, we began by developing a MACE potential[36], which was trained on a comprehensive dataset encompassing crystalline, interfacial, and amorphous configurations (Supplementary Fig. 1). This potential shows high fidelity in energy and force predictions (Supplementary Fig. 2 and Supplementary Table 2) and accurately reproduces the phonon spectra of both 4H-SiC and *β*-Ga$_2$O$_3$ (Supplementary Fig. 3).

Building on this accurate MLIP, we implemented a WPHM grounded in rigorous lattice dynamics (see Methods). This framework overcomes the limitations of classical MD methods by incorporating quantum statistics and capturing phonon coherence, while also bridging the gap between analytical models such as the acoustic and diffuse mismatch models, which respectively assume purely wave-like or particle-like transport. For any incident phonon originating from the first Brillouin zone of *β*-Ga$_2$O$_3$ (Fig. 1a), WPHM enumerates all admissible reflection and transmission channels governed by conservation laws, enabling the simultaneous resolution of mode-resolved phonon transmittance and the spatial evolution of phonon eigenvectors across disordered interfaces.

To quantify the effect of interfacial disorder on phonon transport, we consider a series of *β*-Ga$_2$O$_3$/4H-SiC interfaces with systematically increasing disorder. The disorder-free interface corresponds to an ideal crystalline junction, in which *β*-Ga$_2$O$_3$ forms strong interfacial Si–O bonds with 4H-SiC (Fig. 1b). Increasing disorder was introduced by inserting amorphous transition layers with mixed SiO$_2$–*β*-Ga$_2$O$_3$ compositions and thicknesses ranging from 0 to 4 nm (see Methods). All interface structures were subsequently optimized to thermodynamic stability, and representative geometries are shown in Supplementary Fig. 5. Hereafter, interface models are labelled according to the thickness of the disordered transition layer.

To visualize the microscopic scattering dynamics, we analyzed the spatial evolution of phonon eigenvectors within the interface region. A representative phonon from *β*-Ga$_2$O$_3$ with frequency 3.77 THz, wave vector [0.0, 0.03, −0.04] Å$^{-1}$, and band index 4 was tracked, with atomic displacement profiles shown in Fig. 1b. The simulation system was expanded fourfold in the horizontal direction to clearly visualize the Bloch waves. At the disorder-free (0-nm) interface, the phonon retains a periodic, coherent wave-like character suitable for efficient transmission. In contrast, the introduction of disorder destroys this periodicity, causing phonons to become heavily localized within the disordered interlayer. As a result, the vibrational amplitude transmitted into the 4H-SiC side is strongly attenuated, indicating a substantial suppression of phonon energy transfer.

We quantified this disruption through spatial amplitude and participation ratio (PR) analysis. The resulting average atomic displacement across the interface is shown in Fig. 1c. At the disorder-free (0-nm) interface, the phonon vibrational amplitudes exhibit minimal attenuation, whereas the disordered interfaces strongly suppress vibrational amplitudes. Further thickening of the disordered interlayer does not reduce amplitude, but an anomalous displacement enhancement near the interface indicates pronounced phonon localization induced by the disordered layer. Spatial coherence was assessed via the phonon PR, which approaches 1 for fully extended (coherent) modes and declines toward 0 for localized (incoherent) modes. At the sharp interface, the PR exhibits only a minor drop before returning near unity on the 4H-SiC side, indicating sustained coherent transport (Fig. 1d). Disordered interfaces show markedly lower PR values on both sides, indicating a significant loss of phonon coherence during transmission. Consequently, both transmittance and specularity peak at the disorder-free limit but rapidly decay and saturate as disorder increases (Fig. 1e).

To further clarify the factors governing phonon scattering at disordered interfaces, we examined a set of phonons propagating along the same crystallographic direction but with different wave-vector magnitudes. The corresponding transmitted and reflected states were mapped in reciprocal space (Fig. 1f). At the disorder-free interface, phonon transmittance remains uniformly high across the entire wavelength range. By contrast, disordered interfaces preferentially transmit long-wavelength phonons while strongly suppressing short-wavelength components, indicating that interfacial disorder effectively acts as a low-pass filter for phonon transport.

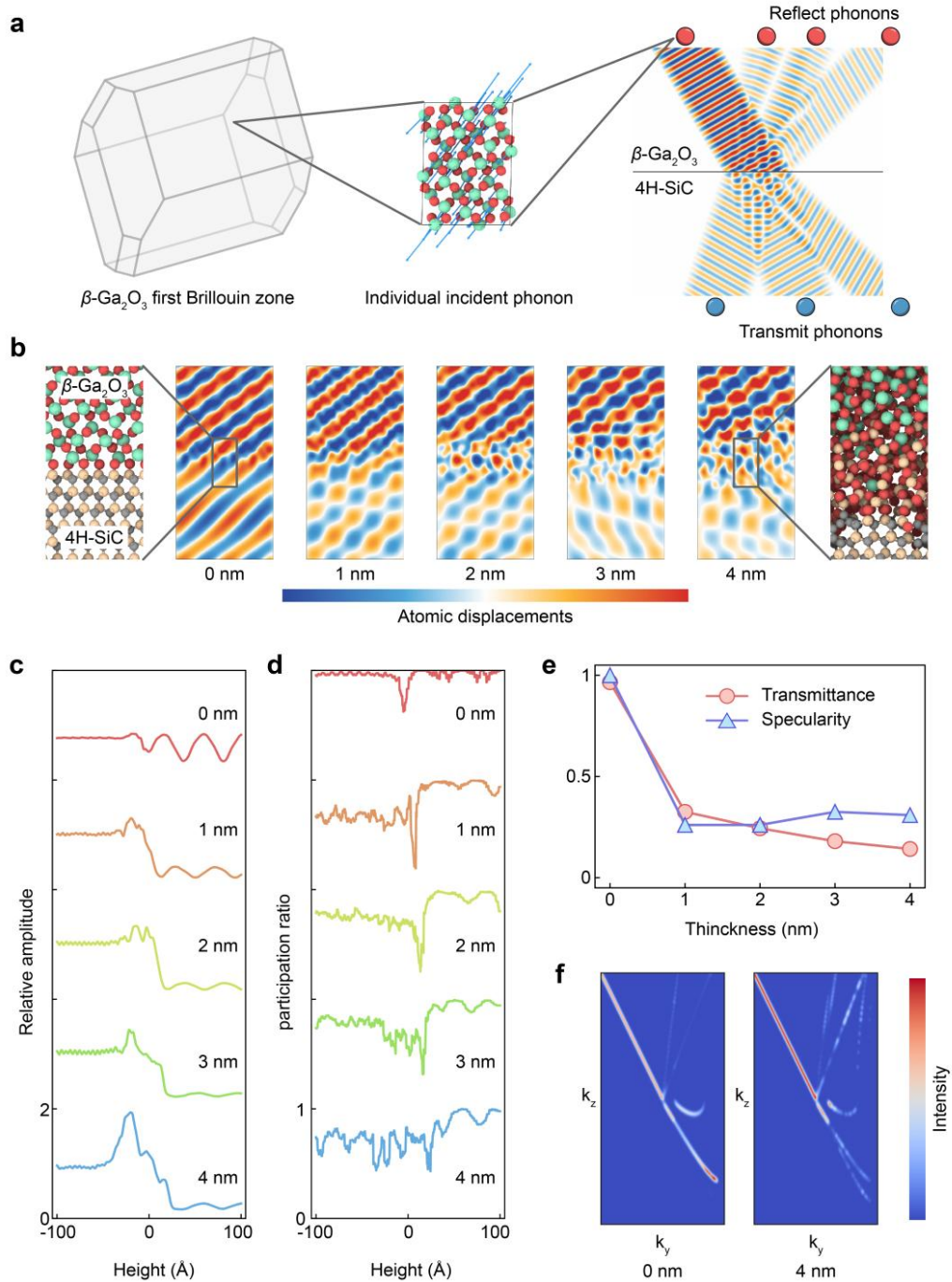

**Fig. 1 | Effect of interfacial disorder on individual phonon transport across $\beta$-Ga$_2$O$_3$/4H-SiC interfaces. a,** Schematic illustration of the simulation framework. **b,** Real-space atomic displacement maps for a representative phonon at interfaces containing amorphous transition layers of different thicknesses (0–4 nm). **c,** Profiles of relative phonon displacement amplitude across the interface, demonstrating strong attenuation of vibrational amplitude with increasing interfacial disorder. **d,** PR distributions across the interface. **e,** Dependence of phonon transmittance and specularity on interfacial disorder thickness. **f,** Reciprocal-space visualization of phonon scattering for a disorder-free interface (0-nm, left) and a disordered interface (4-nm, right).

**Disorder-mediated diffuse scattering and phonon bridging**

Interface disorder introduces a complex interplay between two concurrent mechanisms: enhanced diffuse scattering, which randomizes phonon momentum and redirects transmission, and phonon bridging, where intermediate vibrational states serve to bridge the spectral mismatch. To rigorously quantify these effects, we performed calculations on a uniform 14 × 14 × 14 *k*-point grid spanning the first Brillouin zone. The WPHM identifies all conservation-allowed outgoing channels for each incident phonon mode, enabling transport events to be classified as either specular (conserving in-plane momentum) or diffuse. This capability allows an explicit decomposition of the total heat flux into specular and diffuse components, providing a microscopic classification of transport channels that extends beyond the purely specular and purely diffuse limits assumed in the AMM and DMM, respectively (Figs. 2a–c).

    The distribution of scattered phonon energy flux is presented in Fig. 2d. Regarding specular scattering, specular transmission comprises a significant fraction (15%) of the total heat flux at the disorder-free (0-nm) interface. However, this contribution drops precipitously with the introduction of interfacial disorder; a 1-nm disordered layer suppresses specular transmission to 2.6%, with further reduction observed as the disorder thickness increases. Conversely, specular reflection—initially 60% at the ideal interface—remains relatively robust at disordered interfaces, accounting for 23–29% of the total reflected flux. In terms of diffuse scattering, a substantial portion of energy (approximately 15%) undergoes diffuse reflection even at the ideal interface. Upon the introduction of disorder, this contribution surges to 60% and subsequently stabilizes. Meanwhile, diffuse transmission remains relatively stable, varying from 10% at the disorder-free interface to 8% at the disordered interfaces. Although a sudden shift is noted at the 1-nm interface, this primarily represents a mode conversion from specular to diffuse transmission rather than an intrinsic change in total transmission dynamics. The heat transport is proportional to the combined energy flux of specular and diffuse transmission. Based on WPHM calculations, the disorder-free interface exhibits the highest TBC (215 MW m$^{-2}$ K$^{-1}$), whereas it declines (165–83 MW m$^{-2}$ K$^{-1}$) as the degree of disorder increases. Notably, the AMM, which neglects diffuse scattering, significantly overestimates the TBC (543 MW m$^{-2}$ K$^{-1}$); similarly, the DMM, by ignoring specular scattering, also yields an inflated TBC (388 MW m$^{-2}$ K$^{-1}$). The failure of both AMM and DMM stems fundamentally from their inability to resolve these distinct transport channels. Consequently, explicitly decomposing the heat flux into

these four components is a prerequisite for accurate TBC prediction.

To elucidate the role of the bridging mechanism, we calculated the vibrational density of states (VDOSs) for $β$-Ga$_2$O$_3$, the disordered interfacial layer, and 4H-SiC (Fig. 2e and Supplementary Fig. 6). While $β$-Ga$_2$O$_3$ and 4H-SiC exhibit intrinsic overlap in the 7–15 THz range, the disordered layer actively modifies this connectivity. Quantitative analysis shows that the fractional VDOS overlap increases when the disordered layer is included, following the trend 4H-SiC–disordered > $β$-Ga$_2$O$_3$–disordered > direct $β$-Ga$_2$O$_3$–4H-SiC (Fig. 2f), indicating that the disordered region introduces additional vibrational states that connect the two crystalline spectra. Although the net transmitted flux decreases, this result highlights the nuanced nature of the interface: the enhanced spectral matching provided by the "phonon bridge" coexists with, yet is counterbalanced by, the increased diffuse reflection and momentum randomization intrinsic to the disordered structure.

Frequency-resolved transmittance results further illustrate this balance between scattering and bridging. The transition from the disorder-free interface to the disordered cases shows a monotonic adjustment in transmission levels (Fig. 2g). Crucially, within the 7–15 THz window, the transport is sustained by the interplay of these mechanisms. Specularity spectra verify the microscopic shift: disorder transforms specular phonon events into diffuse ones across the spectrum (Fig. 2h). Consequently, the observed TBC reflects the integration of these two parallel effects: the disorder introduces intermediate modes that successfully bridge the vibrational gap, while simultaneously enhancing diffuse scattering which modulates the overall transmission efficiency. Beyond the spectral analysis, the WPHM also resolves transmission with respect to incident and azimuthal angles[29] (Supplementary Fig. 7). We observe that phonon transmittance decays with increasing incident angle and displays azimuthal anisotropy consistent with the monoclinic symmetry of $β$-Ga$_2$O$_3$.

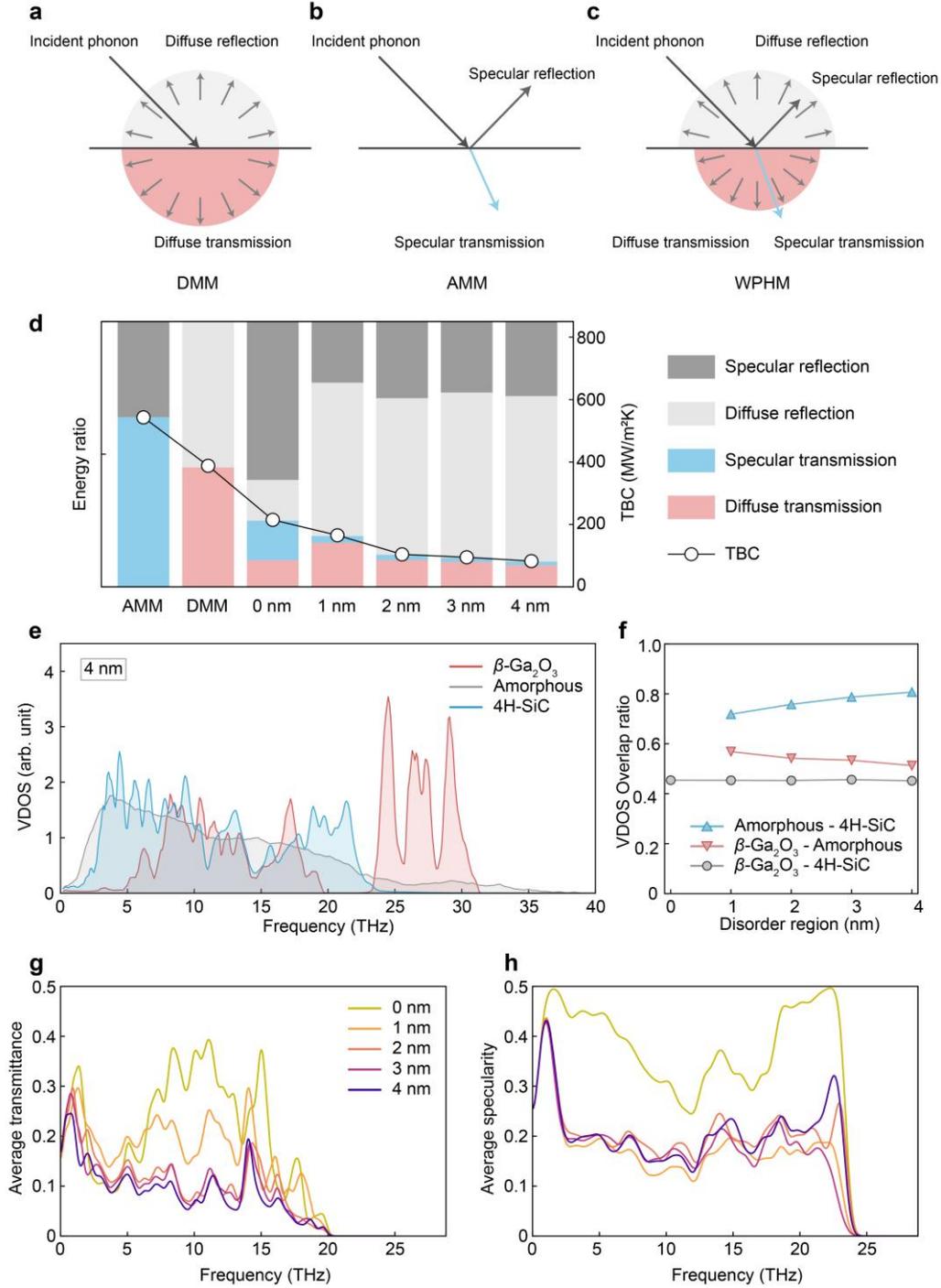

**Fig. 2 | Disorder-induced competition of phonon scattering channels at interfaces.**
**a–c,** Schematics of phonon reflection and transmission under the DMM, AMM, and WPHM. **d,** Energy partitioning into specular reflection, diffuse reflection, specular transmission, and diffuse transmission for different disorder thicknesses, with corresponding thermal boundary conductance. **e,** VDOS of $\beta$-Ga$_2$O$_3$, the amorphous interlayer, and 4H-SiC. **f,** VDOS overlap ratios between material pairs as a function of disorder thickness. **g,** Frequency-dependent average transmittance for interfaces with increasing disorder. **h,** Average specularity as a function of frequency for the same interfaces.

**Epitaxial interfaces and thermal boundary conductance**

Motivated by theoretical predictions that disorder-free interfaces maximize TBC, we pursued an epitaxial route to synthesize $\beta$-Ga$_2$O$_3$/4H-SiC heterostructures. In contrast to the more commonly used wafer-bonding approach, which typically relies on adhesives or intermediate layers that introduce interfacial disorder and obscure intrinsic phonon transport[31,32,35,37–42], we adopted radio-frequency magnetron sputtering under carefully optimized growth conditions (see Methods). This strategy enables the formation of atomically sharp interfaces without bonding-induced interlayers[43,44], providing a platform for experimentally probing the TBC of structurally pristine $\beta$-Ga$_2$O$_3$/4H-SiC interfaces. Achieving high-quality epitaxy remains challenging, however, as the elevated growth temperatures required for $\beta$-Ga$_2$O$_3$ can promote substrate oxidation, necessitating precise control over the deposition process.

To address this challenge, we developed a synthesis protocol that suppresses substrate oxidation and disordered interlayer formation. By carefully controlling the temperature and trace oxygen level in the background vacuum (Fig. 3a, see Methods), we successfully achieved epitaxial growth of $\beta$-Ga$_2$O$_3$ films on 4H-SiC substrates with systematically varied interface structures. Three representative samples were prepared to map this evolution, designated Sample 1 (4-nm disorder), Sample 2 (3-nm disorder), and Sample 3 (atomically sharp, 0-nm). Specifically, the disordered interlayer between $\beta$-Ga$_2$O$_3$ and 4H-SiC begins to form and progressively increases in thickness as both the growth temperature and the trace oxygen level are elevated.

The structural quality and interface morphology of $\beta$-Ga$_2$O$_3$/4H-SiC heterostructures were investigated using X-ray Diffraction (XRD) and high-angle annular dark-field scanning transmission electron microscopy (HAADF-STEM). XRD analysis confirmed that all samples exhibit a single ($\bar{2}$01) orientation, indicating high crystalline quality (see Supplementary Figs. 8–9). Figs. 3b–d and Supplementary Fig. 10 present cross-sectional HAADF-STEM images of the interface regions for Sample 1, Sample 2, and Sample 3, respectively. For Samples 1 and 2 (Figs. 3b–c), three distinct layers are clearly identifiable: the 4H-SiC substrate, the disordered interfacial layer, and the $\beta$-Ga$_2$O$_3$ epitaxial film. As highlighted by the orange dashed lines, the disordered interfacial layer thickness is approximately 4 nm for Sample 1 and 3 nm for Sample 2. XPS analysis reveals that the formation of this layer is primarily driven by the oxidation of the SiC surface under high temperature and residual oxygen, which induces random $\beta$-Ga$_2$O$_3$ nucleation at the initial growth stage (Supplementary Figs. 11–13), and consequently leads to a significant increase in the film's surface roughness (Supplementary Fig. 14). Additionally, atomic interdiffusion at high temperatures further contributes to the formation of this amorphous layer (Supplementary Fig. 15). In marked contrast, Sample 3 shows an atomically sharp interface with no detectable disordered interface (Fig. 3d). The $\beta$-Ga$_2$O$_3$ layer exhibits highly ordered atomic arrangement with a preferred orientation. The rocking curve full width at half maximum (FWHM) for this sample is also notably narrow (0.205°; see XRD, Supplementary Fig. 8b), confirming superior crystallinity. This high-quality interface is particularly crucial for attaining superior TBC according to our theoretical predictions.

The TBC across the synthesized $\beta$-Ga$_2$O$_3$/4H-SiC interfaces was measured using time-domain thermoreflectance (TDTR), and the results are summarized in Fig. 3e together with representative values from the literature[31,39,35,33,34,45–51]. The disorder-free interface (Sample 3) exhibits an exceptional TBC of 231 MW m$^{-2}$ K$^{-1}$. In stark contrast, the introduction of disordered interlayers in Samples 2 and 1 precipitates a significant decline in thermal coupling, reducing TBC to 150 MW m$^{-2}$ K$^{-1}$ (3-nm disorder) and 111 MW m$^{-2}$ K$^{-1}$ (4-nm disorder), respectively. This monotonic degradation provides experimental verification of the "disorder-mediated diffuse scattering" mechanism: the disordered layer acts as a strong scattering center that randomizes phonon momentum and suppresses the high-transmission specular channels critical for efficient heat removal. Measurement uncertainties were rigorously estimated at ±5% for thermal conductivity and ±20% for TBC (Supplementary Fig. 16).

Notably, the measured trend and magnitude of TBC across Samples 1–3 are in quantitative agreement with our theoretical predictions (Fig. 3e), providing a stringent validation of the underlying transport picture in the presence and absence of an interfacial disordered layer. Finite element simulations of a $\beta$-Ga$_2$O$_3$ multi-gate device demonstrate the practical impact of this improvement: increasing the TBC from 111 MW m$^{-2}$ K$^{-1}$ to 231 MW m$^{-2}$ K$^{-1}$ directly lowers the peak channel temperature by 22 K (Supplementary Note 1 and Supplementary Fig. 17). These results confirm that eliminating the parasitic disordered layer is the decisive factor in unlocking the full thermal potential of $\beta$-Ga$_2$O$_3$/4H-SiC heterogeneous integration.

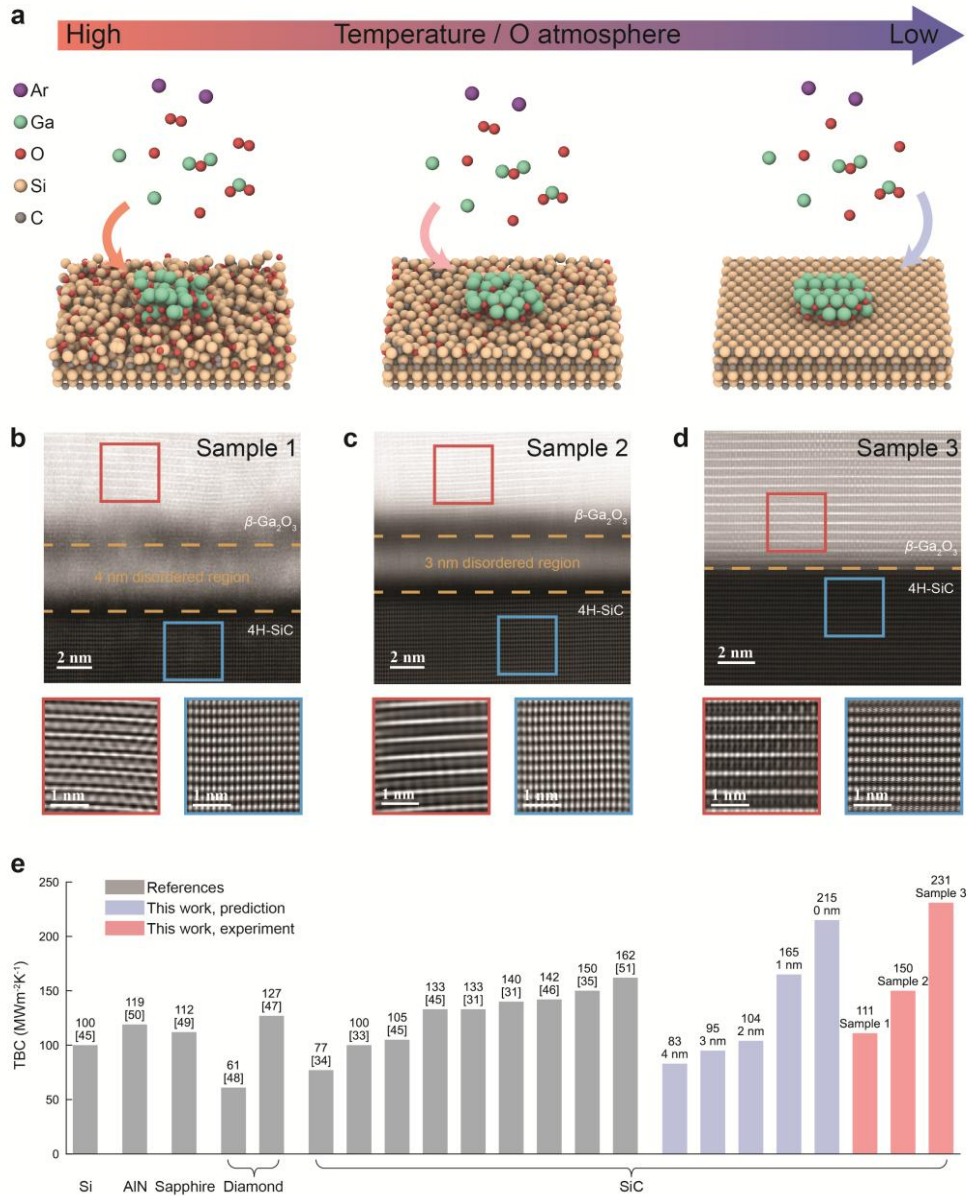

**Fig. 3 | Influence of growth conditions on interfacial structure and TBC of $\beta$-Ga$_2$O$_3$/4H-SiC heterostructures. a,** Schematic illustration showing the evolution of the $\beta$-Ga$_2$O$_3$/4H-SiC interface under varying temperature and residual oxygen levels. Elevated temperatures and higher concentrations promote interfacial oxidation and disorder, while reduced values of both parameters yield a sharp, ordered interface. **b–d,** Cross-sectional STEM images of Samples 1, 2, and 3, respectively; insets show crystalline ordering in both materials. **e,** The measured TBCs for Samples 1–3 (red bars) and theoretical predictions (blue bars) compared with literature data for $\beta$-Ga$_2$O$_3$ on various substrates (gray bars), highlighting the record-high TBC (231 MW m$^{-2}$ K$^{-1}$) achieved in Sample 3.

# Conclusions and Outlook

By integrating machine-learned interatomic potentials with a wave–particle hybrid model, we have resolved the microscopic origins of the thermal bottleneck at $\beta$-$Ga_2O_3$/4H-SiC interfaces. Our theory reveals, from the perspective of phonon eigenvectors, that the primary effect of disorder is the disruption of phonon coherence. Detailed decomposition of the phonon scattering energy flux shows that although disorder induces spectral bridging states, the concomitant loss of specular transmission severely compromises the overall heat flux. This mechanism forces transport into an inefficient diffuse regime, establishing disorder as a primary cause of the thermal bottleneck rather than a coupling aid. Guided by these predictive insights, we experimentally synthesized a series of $\beta$-$Ga_2O_3$/4H-SiC heteroepitaxial junctions with systematically controlled interface quality and quantitatively validated the theoretical picture. The disorder-free interface achieves an experimentally record-high TBC of 231 MW m$^{-2}$ K$^{-1}$, in good agreement with our predicted upper bound for this system.

Beyond its implications for thermal management in $\beta$-$Ga_2O_3$-based ultrawide-bandgap electronics, this work establishes a practical framework for analyzing phonon transport at structurally complex interfaces with experimentally relevant accuracy, overcoming a central barrier to the predictive modelling and design of interfacial thermal transport. Ultimately, while focused on the $\beta$-$Ga_2O_3$/4H-SiC interface, our results imply that maximizing thermal transport in lattice-mismatched systems relies critically on preserving phonon phase coherence through atomically sharp interfaces, rather than relying solely on spectral matching.

# Methods

**Machine-learned interatomic potential**

We developed a MLIP for the $\beta$-$Ga_2O_3$/4H-SiC heterostructure using the MACE framework. To ensure robust transferability across diverse local environments, the training dataset encompassed crystalline $\beta$-$Ga_2O_3$ and 4H-SiC, ideal Si-/O-terminated interfaces, and various amorphous phases ($SiO_2$, $Ga_2O_3$, SiC, the $Ga_2O_3$/SiC interface, and a $Ga_2O_3$–SiC mixture), as shown in Supplementary Fig. 1.

To accelerate dataset generation, we utilized the universal potential MACE-MP-0b3[52] for efficient phase-space sampling, thereby circumventing the high computational cost associated with *ab initio* molecular dynamics. For the crystalline structures ($\beta$-$Ga_2O_3$, 4H-SiC, and the ideal interfaces), we first optimized their lattice constants and then generated nine strained configurations by scaling the lattices from –4% to +4% in 1% increments. Each configuration underwent a 10,000-step (1 fs timestep) MD simulation at 1000 K, driven by MACE-MP-0b3 and regulated by a Langevin thermostat. From each trajectory, we extracted 20 representative structures using the CUR decomposition method, yielding 180 initial configurations per crystalline phase. For the amorphous phases ($SiO_2$, $Ga_2O_3$, SiC, the $Ga_2O_3$/SiC interface, and a $Ga_2O_3$–SiC mixture), we generated 120 random structures using the buildcell program[53]. Each structure was first relaxed to a low-energy state via 5000 steps of L-BFGS optimization[54], followed by a 10,000-step (1 fs timestep) MD simulation at 1000 K using MACE-MP-0b3. After optimization, molecular dynamics simulations at a temperature of 1000 K for 10,000 steps with a timestep of 1 fs were performed, again driven by MACE-MP-0b3. We selected the final configuration from each MD trajectory and subsequently generated compressed variants at 50%, 60%, 70%, 80%, and 90% of the original volume to enhance the potential's robustness under high pressure. All structures in this dataset were then relabeled using density functional theory (DFT) to provide ground-truth energies and forces. The dataset comprises a total of 1,738 cells and 652,240 atoms (Supplementary Table 1). The dataset was then split into 80% training, 10% validation, and 10% test sets. The final MACE model was trained on the training set for 1,000 epochs using a Huber loss function, with two message-passing layers comprising 256 invariant channels and a cutoff radius of 5 Å.

**DFT calculation**

All DFT calculations were conducted using the Vienna Ab-Initio Simulation package[55] with the Perdew–Burke–Ernzerhof exchange–correlation functional[56] and the projector augmented-wave method[57]. An energy cutoff of 600 eV was employed for the plane-wave basis set, and the energy convergence criterion for the electronic self-consistent field iterations was set to $10^{-6}$ eV. Due to the large size of the supercell, the Brillouin zone was sampled using the Γ-point.

**Model construction**

We first constructed a disorder-free interface model by stacking the O-terminated ($\bar{2}01$) β-Ga$_2$O$_3$ surface onto the Si-terminated (001) 4H-SiC surface, followed by structural optimization. To model a disordered interface, we then generated an amorphous SiO$_2$ interlayer by removing SiC from the interfacial region. We then subjected this structure—specifically the amorphous SiO$_2$, a 1-nm underlying 4H-SiC layer, and the adjacent β-Ga$_2$O$_3$ region—to a simulated melt-anneal process. This process involved melting the system at 3,000 K and subsequently annealing it at 883 K to emulate experimental conditions. Finally, we quenched the structure to 0 K and relaxed it to its equilibrium configuration using the L-BFGS algorithm. Supplementary Fig. 5 illustrates the heating-annealing process and the resulting interface structures.

**Lattice dynamics calculation**

To implement the LD method, we divide our system into three distinct regions: a left semi-infinite β-Ga$_2$O$_3$ lead, a central device region containing the β-Ga$_2$O$_3$/4H-SiC interface, and a right semi-infinite 4H-SiC lead. All regions maintain periodicity in the directions parallel to the interface. The eigenvalue equation governs the dynamic behavior of this interfacial system:

$$\begin{pmatrix} H_{LL} & H_{LD} & 0 \\ H_{DL} & H_{DD} & H_{DR} \\ 0 & H_{RD} & H_{RR} \end{pmatrix} \begin{pmatrix} u_L \\ u_D \\ u_R \end{pmatrix} = \omega^2 \begin{pmatrix} u_L \\ u_D \\ u_R \end{pmatrix}, \qquad (1)$$

where $H$ is the dynamical matrix, $\omega$ is the frequency, and $u$ are the eigenvectors. Due to the infinite length of the leads, $u_L$ and $u_R$ are infinite, which makes direct solution impossible. We therefore introduce the following reductions to render the system finite-dimensional. For any incident phonon, we identify all admissible reflected and transmitted phonons by enforcing energy and momentum conservation: the reflected and transmitted phonons share the incident frequency, and their in-plane wave vectors differ from that of the incident phonon by integer multiples of the supercell reciprocal lattice vectors. Then $u_L$ can be expressed as the sum of the incident and all reflected phonon modes, while $u_R$ can be expressed as the sum of all transmitted phonon modes. This formulation yields a linear system with a finite number of variables ($u_D$, $A_{refl}$, and $A_{trans}$),

$$H_{DL}u_L + H_{DD}u_D + H_{DR}u_R = \omega^2 \left( u_L + u_D + u_R \right)$$
$$u_L = u_{inc} + \sum_{refl} A_{refl} u_{refl} \qquad (2)$$
$$u_R = \sum_{trans} A_{trans} u_{trans}$$

The seemingly complicated equation is a linear system and can be rewritten in standard matrix form,

$$A_1 x_1 + A_2 x_2 = b \qquad (3)$$

where $A_1$, $A_2$ and $b$ are coefficient matrices. $x_1$ and $x_2$ are variable arrays, defined by

$$x_1 = \{u_i, i \in D\}$$
$$x_2 = \left\{ A_{R_1}, ..., A_{R_{jmax}}, A_{T_1}, ..., A_{T_{kmax}} \right\} \qquad (4)$$

The energy conservation constraint is formulated as:

$$\mathcal{T} = \sum_{T_k} \frac{v_{T_k,z}}{v_{I,z}} | A_{T_k} |^2,$$
$$\mathcal{R} = \sum_{R_j}^{\infty} \frac{-v_{R_j,z}}{v_{I,z}} | A_{R_j} |^2, \qquad (5)$$
$$\mathcal{T} + \mathcal{R} = 1$$

where $\mathcal{T}$ is transmittance and $\mathcal{R}$ is reflectance. With an appropriate linear rescaling, the energy-conservation constraint can be reduced to $\|x_2\|_2 = 1$.

As the system is overdetermined, we seek an approximate solution by first performing a QR decomposition on $A_1$,

$$A_1 = QR, \qquad (6)$$

where $Q$ is an orthogonal matrix and $R$ is an upper triangular matrix. Subsequently, we optimize $x_1$ and $x_2$ sequentially to minimize the residual, subject to the energy conservation constraint:

$$\begin{aligned}
r_{min} &= \min_{x_2} \min_{x_1} |(b - A_2 x_2) - A_1 x_1| \\
&= \min_{x_2} |(b - A_2 x_2) - QRR^{-1}Q^*(b - A_2 x_2)| \\
&= \min_{x_2} |(I - QQ^*)b - (I - QQ^*)A_2 x_2| \\
&= \min_{x_2} |b_{eff} - A_{eff} x_2|, \\
b_{eff} &= (I - QQ^*)b, \\
A_{eff} &= (I - QQ^*)A_2.
\end{aligned} \tag{7}$$

To solve this constrained minimization problem, we set up the Lagrangian function:

$$L(x_2, \lambda) = \frac{1}{2}\|A_{eff} x_2 - b_{eff}\|_2^2 + \frac{1}{2}\lambda(\|x_2\|_2^2 - 1) \tag{8}$$

$\partial L(x_2, \lambda)/\partial x_2 = 0$ yields:

$$A_{eff}^*(A_{eff} x_2 - b_{eff}) + \lambda x_2 = 0 \tag{9}$$

Perform SVD decomposition on $A_{eff}$:

$$A_{eff} = U\Sigma V^* \tag{10}$$

Then Eq.9 can be rewritten as:

$$\begin{aligned}
(\Sigma^2 + \lambda I) y &= d \\
y &= V^* x_2 \\
d &= \Sigma U^* b_{eff}
\end{aligned} \tag{11}$$

The coefficient matrix $(\Sigma^2 + \lambda I)$ is diagonal, so the solution of $y$ is easy to compute:

$$y_i = \frac{\sigma_i d_i}{\sigma_i^2 + \lambda} \tag{12}$$

The constraint can be written as a scalar equation of $\lambda$:

$$\begin{aligned}
g(\lambda) &= \|x_2\|_2^2 - 1 \\
&= \|y\|_2^2 - 1 \\
&= \sum_i \frac{\sigma_i d_i}{\sigma_i^2 + \lambda} - 1 \\
&= 0
\end{aligned} \tag{13}$$

which is strictly decreasing on its domain $[\sigma_{min}^2, \infty)$. Therefore, $\lambda$ can be readily found by Newton's method. Substituting $\lambda$ back into Eq.11 yields $x_2$ directly. The transmittance of individual phonon mode is then computed with Eq.5.

By sampling phonons on a uniform grid within the first Brillouin zone and computing the transmittance of each phonon, the total TBC is given by the Landauer formula:

$$\text{TBC} = \frac{1}{V} \sum_{qv} \hbar \omega_{qv} v_{qv,z} \mathcal{T}_{qv} \frac{\partial f}{\partial T} \tag{14}$$

where $V$ is the unit cell volume, $\omega_{qv}$ represents the phonon energy at wavevector $q$ and branch $v$, $\partial f / \partial T$ is the derivative of the Bose-Einstein distribution, $v_{qv,z}$ denotes the phonon group velocity component normal to the interface, and $\mathcal{T}_{qv}$ is transmittance.

**Material synthesis**

The $\beta$-$Ga_2O_3$/4H-SiC heterojunction was synthesized via the radio frequency magnetron sputtering on commercially available Si-terminated 4H-SiC substrate. A two-inch $Ga_2O_3$ target was fabricated by conventional solid-state reaction, involving pressing high-purity $Ga_2O_3$ powder (99.99%) followed by sintering at 1450℃ for 8 h in air. For the fabrication of samples with sharply atomic interface (sample 3), the deposition chamber was first evacuated to a base pressure below $4×10^{-4}$ Pa. Prior to heating, high-purity argon (99.999%) was introduced to displace residual atmospheric gases, thereby mitigating substrate surface oxidation during thermal processing. The system was then stabilized under an argon atmosphere at 1.2 Pa, followed by ramping the temperature at a controlled rate of 10 °C/min. The substrate was subjected to a 20-minute in-situ annealing step prior to deposition. Film growth was conducted at 610 °C under a 24 sccm argon flow and a working pressure of 1.2 Pa, with an RF power of 80 W.

For samples incorporating amorphous interlayers (sample 1 and 2), the chamber was evacuated to a higher base pressure of approximately $1×10^{-3}$ Pa, which retained trace oxygen levels. The substrate temperature was then raised directly to the target values—630 °C for sample 2 (3 nm interlayer) and 660 °C for sample 1 (4 nm interlayer) —at a heating rate of 10 °C/min. A 20-minute in-situ annealing pretreatment was also applied to the substrate before deposition under otherwise identical conditions. Subsequent to growth, all samples underwent an additional 30-minute in-situ annealing process to enhance crystallization quality.

**X-ray diffraction**

X-ray diffraction (XRD) measurements were performed using a 9 kW Rigaku diffractometer equipped with a rotating Cu anode producing Cu-Kα radiation (λ = 1.5405 Å). The epitaxial quality and crystallographic orientation of the β-Ga$_2$O$_3$ films were characterized by ω–2θ scans, rocking curves, and φ-scans. The ω–2θ scans were performed over an angular range of 10°–80° to assess crystalline quality and film orientation. Rocking curves of the ($\bar{2}$01) reflection were measured to evaluate the out-of-plane mosaic spread. Phi-scans of the β-Ga$_2$O$_3$ (111) and 4H-SiC (101) planes were acquired to determine the in-plane crystallographic orientation relationship. The inclination angle between the β-Ga$_2$O$_3$ ($\bar{2}$01) and (111) planes is 79.60°, whereas the angle between 4H-SiC (001) and (101) planes is 47.05°. For the β-Ga$_2$O$_3$ (111) φ-scan, 2θ and tilt angle were set as 18.95° and 79.60°, respectively, for β-Ga$_2$O$_3$ (111) planes. For 4H-SiC (101) planes, these values were set as 35.68° and 47.05°, respectively. Both φ-scan profiles were obtained without remounting the sample.

**Electron microscopy**

Cross-sectional transmission electron microscopy (TEM) specimens were prepared using a focused ion beam (FIB) system from three different locations on each sample. Prior to FIB milling, the samples were protected with Au and Pt beam-ionization protective layers. High-angle annular dark-field scanning transmission electron microscopy (HAADF-STEM) images were acquired using an FEI Titan Themis Z microscope (300 kV) equipped with double correctors. Fast Fourier transform (FFT) and inverse FFT analyses were employed to index the reflecting planes and determine the crystallographic orientation relationships.

**Atomic force microscopy**

Surface topography of the β-Ga$_2$O$_3$/4H-SiC heterostructures was characterized in tapping mode using a Bruker Dimension Icon atomic force microscopy equipped with a Nanoscope V controller.

**TDTR measurement**

Thermal properties were characterized using TDTR[58]. A titanium-sapphire laser system (Coherent Mira 900) operating at 781 nm wavelength was employed. The laser output was split into pump and probe beams with powers of 80 mW and 8 mW respectively. The pump beam was modulated at either 10.7 MHz or 1.49 MHz using an electro-optic modulator, while the probe beam was chopped at 200 Hz. Measurements were performed using both 5× and 10× objective lenses, yielding 1/e2 spot radii of 10.13 μm and 5.25 μm respectively, as determined by spatial beam profiling. An 80 nm Pd transducer layer was deposited via electron beam evaporation. The precise Pd thickness was determined from the acoustic echo in the first 100 ps of the TDTR signal. The thermal response was modeled using a multilayer heat diffusion equation in cylindrical coordinates. The 4H-SiC substrate was treated as semi-infinite with known thermal conductivity (270 W m$^{-1}$ K$^{-1}$) and heat capacity (2.28×10$^6$ J m$^{-3}$ K$^{-1}$). Known volumetric heat capacities were used for Pd (2.87×10$^6$ J m$^{-3}$ K$^{-1}$) and $β$-Ga$_2$O$_3$ (2.9×10$^6$ J m$^{-3}$ K$^{-1}$). The $β$-Ga$_2$O$_3$/4H-SiC interface was modeled as a 1 nm layer with minimal heat capacity. High-frequency (10.7 MHz) measurements were first used to determine the $β$-Ga$_2$O$_3$ thermal conductivity, followed by low-frequency (1.49 MHz) measurements to extract the interface thermal conductance.

## Author contributions

H.Y., Y.L., B.C., Y.Y., and Z.W. conceived the project and designed the study; H.Y. performed the main calculations with input from Y.L., B.C., and G.C.; Y.Y. prepared the samples and conducted sample characterization; Y.L. and G.C. developed the MLIP; T.D. performed the TDTR measurements under the supervision of W.M.; Y.S. conducted the FEM simulations; J.H. carried out the TEM characterization under the supervision of L.F.; B.C. supervised the work. H.Y., Y.L., and Y.Y. wrote the original draft. Z.W., G.C., and B.C. reviewed and edited the manuscript. All authors approved the final version. H.Y., Y.Y., and Y.L. contributed equally to this work.

## Data and materials availability

Data supporting this work, including the parameter files required to use the potential, the fitting data, and the structural models, will be made openly available in a suitable repository upon journal publication.

## Acknowledgements

This work was supported by the National Natural Science Foundation of China (Grant Nos. 52425601, 52327809, 52206098, and 12474065), the National Key Research and Development Program of China (Grant No. 2023YFB4404104), the Beijing Natural Science Foundation (No. L233022), and the Fund of the State Key Laboratory of Information Photonics and Optical Communications (Grant Nos. IPOC2025ZR05 and IPOC2025ZJ06).

## Competing interests

The authors declare no competing interests.